\documentclass{article}
\usepackage{bm}
\begin{document}

\title{Constraining Neutrino Mass Matrix from Modified BM with Softly Broken $\mu-\tau$ Symmetry} 
\author{\bf{Asan Damanik}\footnote{E-mal: d.asan@lycos.com}\\ {\it Faculty of Science and Technology, Sanata Dharma University}\\ {\it Kampus III USD Paingan Maguwoharjo Sleman Yogyakarta, Indonesia}}
\date{}

\maketitle

\abstract{The bimaximal (BM) neutrino mixing matrix was formulated in order to accommodate the data of the experimental results which indicate that both solar and atmospheric neutrino oscillation in vacuum are near maximal.  But, after the T2K and Daya Bay Collaborations reported that the mixing angle $\theta_{13}$ is nonzero and relatively large, many authors have modified the neutrino mixing matrix in order to accommodate experimental data.  We modified the BM mixing matrix by introducing a simple perturbation matrix into BM mixing matrix.  The modified BM mixing matrix can proceed the mixing angles which are compatible with the globat fit analysis data and by imposing the $\mu-\tau$ symmetry into mass matrix from modified BM, we have the neutrino mass in normal hierarchy: $m_{1}<m_{2}<m_{3}$.  Using the neutrino masses that obtained from neutrino mass matrix in the scheme of modified BM and imposing the constraint exact $\mu-\tau$ symmetry into neutrino mass matrix, we cannot have compatible squared-mass differences for both $\Delta m_{21}^{2}$ and $\Delta m_{32}^{2}$ as dictated by experimental results.  We break  softly the $\mu-\tau$ symmetry by introducing a small parameter $\lambda$ into neutrino mass matrix which then can proceed neutrino masses are in agreement with the squared mass difference as dictated by experimental results.The predicted neutrino effective mass: $\left|m_{ee}\right|=0.0155~{\rm eV}$ in this paper can be tested in the future neutrinoless double beta decay.}

\section{Introduction}
Neutrino oscillation phenomena can be explained if neutrino has a nonzero  mass and some mixing angles exist in neutrino sector.  The mixing angles for three neutrinos can  be formulated in mixing matrix.  The mixing matrix relate the neutrino eigenstates in flavor basis ($\nu_{e},\nu_{\mu},\nu_{\tau}$) into neutrino eigenstate in mass basis ($m_{1},m_{2},m_{3}$) as follow:
\begin{eqnarray}
\bordermatrix{&\cr
&\nu_{e}\cr
&\nu_{\mu}\cr
&\nu_{\tau}}=V\bordermatrix{&\cr
&m_{1}\cr
&m_{2}\cr
&m_{3}},\label{1}
\end{eqnarray}
where $V$ is the mixing matrix.  The  standard parameterization of the mixing matrix read:
\begin{eqnarray}
V=\bordermatrix{& & &\cr
&c_{12}c_{13} &s_{12}c_{13} &s_{13}e^{-i\delta}\cr
&-s_{12}c_{23}-c_{12}s_{23}e^{i\delta} &c_{12}c_{23}-s_{12} s_{23}e^{i\delta}&s_{23}c_{13}\cr
&s_{12}s_{23}-c_{12}c_{23}e^{i\delta} &-c_{12}s_{23}-s_{12}c_{23}e^{i\delta} &c_{23}c_{13}}
 \label{V}
\end{eqnarray}
where $c_{ij}$ is the $\cos\theta_{ij}$, $s_{ij}$ is the $\sin\theta_{ij}$, $\theta_{ij}$ are the mixing angles, and $\delta$ is the Dirac CP-violating phase.

There are three well-known of neutrino mixing matrix, i.e. bimaximal mixing (BM), tribimaximal mixing (TBM), and democratic mixing (DM).  All of the mixing matrices predict the mixing angle $\theta_{13}=0$. Recently, the experimental results showed that the mixing angle $\theta_{13}\neq 0$ and relatively large \cite{T2K,MINOS,Double,Daya,RENO}.   Several attempt have been done theoretically to accommodate the nonzero and relatively large mixing angle $\theta_{13}$  by modifying the neutrino mixing matrix including the Dirac phase $\delta$ in relation to the CP-violation in neutrino sector.  Another unsolved problem in neutrino physics till today is the hierarchy of neutrino mass.  Experimental results showed that we have two possibilities for neutrino mass hierarchies: normal and inverted hierarchies.  We have no clue in order to decide theoretically the neutrino mass hierarchy whether it normal or inverted.

Theoretically, the neutrino masses can be determined from neutrino mass matrix ($M_{\nu}$).  The neutrino mass matrix  is related to the neutrino mixing matrix via the following equation:
\begin{eqnarray}
M_{\nu}=VMV^{T},\label{3}
\end{eqnarray}
where $M$ is the neutrino mass matrix in mass basis and $V$ is the mixing matrix.  From Eq. (\ref{3}), we can see that the neutrino mass pattern in flavor basis depend on the pattern of the neutrino mixing matrix.  We also belief that the pattern of the neutrino mass matrix should be related to one of the unique underlying symmetry.  One of the interesting candidate symmetry is the $\mu-\tau$ symmetry because it can reduce the number of parameters in neutrino mass matrix and can predicts qualitatively the neutrino mass hierarchy that compatible with the experimental results.  But, if we use the advantage of the experimental results as input to fix the values of some parameters that we build from theoretical side with exact $\mu-\tau$ symmetry, many theoretical predictions are incompatible with the experimental results especially mixing angle $\theta_{13}$ which lead to be zero in the frame of $\mu-\tau$ symmetry.  Thus, if we still want to use the $\mu-\tau$ symmetry, the we should invoke a perturbation into neutrino mass matrix or into three well-known mixing matrices BM, TBM, and DC.  The idea of introducing a perturbation into $\mu-\tau$ symmetry mass matrix have been introduced in \cite{Liao,Gupta,Adhikary,Lashin}, where authors analyzed the effect of perturbation and its correlation corresponding to the mixing angles: $\theta_{12}$ and $\theta_{23}$.  In \cite {AD} the broken $\mu-\tau$ symmetry is used to obtain nonzero $\theta_{13}$ and Jarlskog rephasing invariant.

In this paper, we determine the neutrino mass hierarchy and neutrino masses from the neutrino mass matrix that obtained from the modified neutrino mixing BM with additional constraint softly broken $\mu-\tau$ symmetry as the underlying symmetry of the resulted neutrino mass matrix.  In section II, we modified BM by introducing a simple perturbation matrix and calculate mixing angles $\theta_{12}$ and $\theta_{23}$ by taking the advantage of Daya Bay Collaboration result on mixing angle $\theta_{13}$ \cite{Daya}.  In section III, we evaluate the neutrino mass hierarchy and neutrino masses from the neutrino mass matrix obtained from modified BM with assumption that the underlying symmetry of the neutrino mass matrix is the $\mu-\tau$ symmetry.  We also discuss predictions of the obtained neutrino masses especially on the squared mass difference for both solar and atmospheric neutrinos.  Finally, section IV is devoted for conclusions.

\section{Modified BM Mixing Matrix}
The bimaximal (BM) mixing matrix was formulated in order to accommodate the facts that both solar and atmospheric data can be described by maximal mixing vacuum oscillation with the relevant mass scale and it imply that there is a unique mixing matrix  which is then called bimaximal mixing matrix.  The BM mixing matrix ($V_{BM}$ ) read \cite{Pakvasa}:
\begin{eqnarray}
V_{BM}=\bordermatrix{& & &\cr
&\sqrt{\frac{1}{2}} &-\sqrt{\frac{1}{2}} &0\cr
&\frac{1}{2} &\frac{1}{2} &-\sqrt{\frac{1}{2}}\cr
&\frac{1}{2} &\frac{1}{2} &\sqrt{\frac{1}{2}}}.
 \label{4}
\end{eqnarray}

As one can see from Eq. (\ref{4}), by comparing it to neutrino mixing matrix in Eq. (\ref{V}), the BM mixing matrix give: $\sin\theta_{13}e^{-i\delta}=0$  which imply that the BM mixing matrix leads to mixing angle $\theta_{13}=0$ which is incompatible with the recent experimental results of T2K Collaboration \cite {T2K}:
\begin{eqnarray}
5^{o}\leq \theta_{13}\leq 16^{o},\label{5}
\end{eqnarray}
for neutrino mass in normal hierarchy (NH), and
\begin{eqnarray}
5.8^{o}\leq \theta_{13}\leq 17.8^{o},\label{6}
\end{eqnarray}              
for inverted hierarchy (IH) with Dirac phase: $\delta=0$.  The nonzero value of mixing angle $\theta_{13}$ was also confirmed by Daya Bay Collaboration as follow \cite{Daya}:
\begin{eqnarray}
\sin^{2}2\theta_{13}=0.092\pm 0.016~ ({\rm stat.})\pm 0.005~({\rm syst}).\label{7}
\end{eqnarray}
From Eq. (\ref{4}) we also see that the mixing angle:$\theta_{12}=\theta_{23}=\pi/4$ (bimaximal mixing).
   
In order to accommodate the relatively large and  nonzero mixing angle $\theta_{13}$ in accordance with BM mixing matrix, many authors have modified the BM mixing matrix \cite{Asan,Chao,Garg}.  In this paper we modify the neutrino matrix mixing BM by introducing a simple perturbation matrix to perturb BM which different with the Refs. \cite{Asan,Chao,Garg}.  The perturbation matrix ($V_{p}$) is given by (with Dirac phase: $\delta=0$):
\begin{eqnarray}
V_{p}=\bordermatrix{& & &\cr
&c_{p} &0 & s_{p}\cr
&0 &1 &0\cr
&-s_{p} &0 &c_{p}},\label{8}
\end{eqnarray}
where $c_{p}$ is $\cos{p}$, and $s_{p}$ is $\sin{p}$.  The modified BM mixing matrix ($U_{BM}$) is obtained via the following relation:
\begin{eqnarray}
U_{BM}=V_{p}V_{BM},\label{9}
\end{eqnarray}
which then gives:
\begin{eqnarray}
U_{BM}=\bordermatrix{& & &\cr
&\frac{s_{p}+\sqrt{2}c_{p}}{2} &\frac{s_{p}-\sqrt{2}c_{p}}{2} &\frac{\sqrt{2}s_{p}}{2}\cr
&\frac{1}{2} &\frac{1}{2} &-\frac{\sqrt{2}}{2}\cr
&\frac{c_{p}-\sqrt{2}s_{p}}{2} &\frac{c_{p}+\sqrt{2}s_{p}}{2} &\frac{\sqrt{2}c_{p}}{2}}.\label{10}
\end{eqnarray}

From Eq. (\ref{10}) one can see that the modified BM can predicts nonzero mixing angle $\theta_{13}$.  By comparing Eq. (\ref{10}) with Eq. (\ref{V}), we have:
\begin{eqnarray}
\tan\theta_{12}=\left|\frac{s_{p}-\sqrt{2}c_{p}}{s_{p}+\sqrt{2}c_{p}}\right|,\label{11}\\
\tan\theta_{23}=\left|\frac{1}{c_{p}}\right|,\label{12}\\
\sin\theta_{13}=\left|\frac{\sqrt{2}s_{p}}{2}\right|.\label{13}
\end{eqnarray}
By inspecting Eqs. (\ref{11}), (\ref{12}), and (\ref{13}), we can see that the mixing angles: $\theta_{13}\neq 0$, $\theta_{12}<\pi/4$, and $\theta_{23}>\pi/4$.  On the other hand, the modified BM in this scenario proceed  {\it no-maximal}  mixing matrix.

If we insert the central value of mixing angle $\theta_{13}$ as reported by Daya Bay Collaboration as written in Eq. (\ref{7}), then we have: $p=12.54^{o}$  and it implies mixing angles:
\begin{eqnarray}
\theta_{12}=36.07^{o}~~{\rm and}~~\theta_{23}=45.69^{o},\label{14}
\end{eqnarray}
which is compatible with the global fit analysis data \cite{Gonzales-Garcia}:
\begin{eqnarray}
\theta_{12}=34.5\pm 1.0 (^{+3.2}_{-2.8})^{o},\label{15}\\
\theta_{23}=42.8^{+5.5}_{-2.9}(^{+10.7}_{-7.3})^{o},\label{16}
\end{eqnarray}
for $1\sigma$ ($3\sigma$) level.

\section{Neutrino Mass Hierarchy from $U_{BM}$}
As stated in section I, one of the the unsolved  problem in neutrino physics is the neutrino mass hierarchy whether it normal or inverted.  By using the Eq. (\ref{3}) with neutrino mixing is the modified BM ($U_{BM}$) with mixing angles as shown in Eq. (\ref{14})  and mass matrix ($M$)  in mass basis is diagonal:
\begin{eqnarray}
M=\bordermatrix{& & &\cr
&m_{1} &0 &0\cr
&0 &m_{2} &0\cr
&0 &0 &m_{3}},\label{17}
\end{eqnarray}
then we have the neutrino mass matrix as follow:
\begin{eqnarray}
M_{\nu}=\bordermatrix{& & &\cr
&P &Q &R\cr
&Q &S &T\cr
&R &T &W},\label{18}
\end{eqnarray}
where:
\begin{eqnarray}
P=0.638m_{1}+0.338m_{2}+0.024m_{3},\label{19}\\
Q=-0.904m_{1}+0.084m_{2}+0.108m_{3},\label{20}\\
R=-0.115m_{1}-0.5764m_{2}+0.106m_{3},\label{21}\\
S=0.979m_{1}+0.021m_{2}+0.499m_{3},\label{22}\\
T=-0.143m_{1}+0.143m_{2}+0.488m_{3},\label{23}\\
W=0.021m_{1}+0.979m_{2}+0.477m_{3}.\label{24}
\end{eqnarray}

If we impose the exact $\mu-\tau$ symmetry into resulted neutrino mass matrix ($M_{\nu}$) of Eq. (18), then we must put $Q=R$ or
\begin{eqnarray}
Q-R=0,\label{25}
\end{eqnarray}
and $S=W$ or
\begin{eqnarray}
S-W=0.\label{26}
\end{eqnarray}
Solving simultaneously Eqs. (\ref{25}) and (\ref{26}), we have neutrino masses relations as follow:
\begin{eqnarray}
m_{2}=1.1727m_{1 }~~{\rm and}~~m_{3}=7.5190m_{1},\label{27}
\end{eqnarray}
which imply that:
\begin{eqnarray}
m_{1}<m_{2}<m_{3}.\label{28}
\end{eqnarray}
The resulted hierarchy of neutrino mass in Eq. (\ref{28}) is normal hierarchy (NH).

By taking the advantage of global fit analysis of the experimental results of the squared-mass difference \cite{Gonzales-Garcia,Fogli}:
\begin{equation}
\Delta m_{21}^{2}=7.59\pm0.20 (_{-0.69}^{+0.61}) \times 10^{-5}~\rm{eV^{2}},\label{29}
\end{equation}
\begin{equation}
\Delta m_{32}^{2}=2.46\pm0.12(\pm0.37) \times 10^{-3}~\rm{eV^{2}},~\rm(for~ NH),\label{30}
\end{equation}
then we have:
\begin{eqnarray}
m_{1}=0.014224~{\rm eV},\nonumber\\
m_{2}=0.016680~{\rm eV}, \label{31}\\
m_{3}=0.106949~{\rm eV},\nonumber
\end{eqnarray}
when we use the squared-mass difference of Eq. (\ref{29}) to determine the neutrino mass in Eq. (\ref{27}).  Conversely, if we use the squared-mass difference of Eq. (\ref{30}) to determine the neutrino mass in Eq. (\ref{27}), then we have:
\begin{eqnarray}
m_{1}=0.006678~{\rm eV},\nonumber\\
m_{2}=0.007831~{\rm eV}, \label{32}\\
m_{3}=0.050212~{\rm eV}.\nonumber
\end{eqnarray}

From Eq. (\ref{31}), one can see that the neutrino masses predict the squared-mass difference $\Delta m^{2}_{32}=11.16\times 10^{-3}~{\rm eV^{2}}$  which is incompatible with the experimental result.  It is also apparent from Eq. (\ref{32}) that neutrino masses predict $\Delta m^{2}_{21}=1.673\times 10^{-5}~{\rm eV^{2}}$  which is incompatible with the experimental results.  Thus, in the context of exact  $\mu-\tau$ symmetry as an underlying symmetry of neutrino mass matrix, neutrino masses cannot proceed compatible predictions for both squared-mass differences:$\Delta m^{2}_{21}$ and $\Delta m^{2}_{32}$ as dictated by experimental results.  In order to get the compatible predictions of the obtained neutrino masses with the experimental results, we break softly the $\mu-\tau$ symmetry with the following scenario: $Q=R$ or
\begin{eqnarray}
Q-R=0,\label{33}
\end{eqnarray}
and $S-W=\lambda$ or
\begin{eqnarray}
S-W-\lambda=0,\label{34}
\end{eqnarray}
where $\lambda$  is a small parameter that perturb softly the exact $\mu-\tau$ symmetry.  Solving Eqs. (\ref{33}) and (\ref{34}) simultaneously, we have:
\begin{eqnarray}
m_{1}=0.1329967m_{3}-5.3405836\lambda,\label{35}\\
m_{2}=0.1559612m_{3}-6.3844249\lambda.\label{36}
\end{eqnarray}

Using the neutrino mass relations in Eqs. (\ref{35}) and (\ref{36}) and we equate it with the squared-mass difference of Eq. (\ref{29}) to determine the neutrino mass $m_{3}$, then we have:
\begin{eqnarray}
m_{3}=43.0157073\lambda+0.964468774\times10^{-15}\sqrt{0.63942498\lambda^{2}+0.122962768\times 10^{29}}.\label{37}
\end{eqnarray}
Inserting the global fit analysis result in Eq. (30) into squared-mass difference $\Delta m^{2}_{32}$ which is calculated from Eqs. (\ref{36}) and (\ref{37}), we have the value of parameter $\lambda$ as follow:
\begin{eqnarray}
\lambda=-0.00127381~~{\rm or}~~\lambda=-0.00370340. \label{38}
\end{eqnarray}

By inserting the value of $\lambda=-0.00127361$ into Eqs. (\ref{35}), (\ref{36}), and (\ref{37}), we finally have neutrino masses:
\begin{eqnarray}
m_{1}=0.0137452~{\rm eV},\nonumber\\
m_{2}=0.0162737~{\rm eV}, \label{39}\\
m_{3}=0.0521999~{\rm eV}.\nonumber
\end{eqnarray}
that can proceed the squared-mass differences $\Delta m^{2}_{21}$ and $\Delta m^{2}_{32}$ are compatible with the experimental results.  By using the relation of neutrino effective mass of neurinoless double beta decay:
\begin{eqnarray}
\left|m_{ee}\right|=\left|(U_{BM})^{2}_{e1}m_{1}+(U_{BM})^{2}_{e2}m_{2}+(U_{BM})^{2}_{e3}m_{3}\right|,
\end{eqnarray}
then we have:
\begin{eqnarray}
\left|m_{ee}\right|=0.0155~{\rm eV},
\end{eqnarray}
that can be tested in the future neutrinoless double beta decay experiment.

\section{Conclusions}
We have modified BM mixing matrix by introducing a simple perturbation matrix into BM that can proceed nonzero mixing angle: $\theta_{13}$ and relatively large.  From the modified BM mixing matrix and using the central value of mixing angle: $\theta_{13}$ from Daya Bay Collaboration data, we determine the mixing angles: $\theta_{12}$ and $\theta_{23}$ which are compatible with the global fit analysis data.  The neutrino mass matrix obtained from the modified BM mixing matrix with is constrained by exact  symmetry  proceed the neutrino mass in normal hierarchy: $m_{1}<m_{2}<m_{3}$  , but neutrino masses cannot predict the squared-mass differences: $\Delta m^{2}_{21}$ and $\Delta m^{2}_{32}$ as dictated bay the experimental results.  To get the neutrino masses that can proceed the compatible squared-mass differences: $\Delta m^{2}_{21}$ and $\Delta m^{2}_{32}$  with the experimental results, we break softly the $\mu-\tau$ symmetry by introducing a small parameter $\lambda$ into neutrino mass matrix.  The obtained neutrino masses from softly broken $\mu-\tau$ symmetry as the underlying symmetry of neutrino mass matrix can proceed the squared-mass differences: $\Delta m^{2}_{21}$ and $\Delta m^{2}_{32}$ which are in agreement with the experimental results with neutrino masses: $m_{1}=0.0137452~{\rm eV}$, $m_{2}=0.0162737~{\rm eV}$, and $m_{3}=0.0521999~{\rm eV}$.  The predicted neutrino effective mass: $\left|m_{ee}\right|=0.0155~{\rm eV}$ in this paper can be tested in the future neutrinoless double beta decay

\end{document}